\documentclass[review]{elsarticle}

\usepackage{lineno,hyperref}
 \usepackage{graphicx}
 \usepackage[normalem]{ulem}
\usepackage{amsmath}
\usepackage{amssymb}
\usepackage{color}
\usepackage{verbatim}

\journal{arXiv}

\biboptions{authoryear}

\begin{document}

\begin{frontmatter}

\title{On the scale dependence in the dynamics of frictional rupture: constant fracture energy versus size-dependent breakdown work}


\author[1]{Federica Paglialunga\corref{cor1}\fnref{fn1}}
\cortext[mycorrespondingauthor]{Corresponding author}
\ead{federica.paglialunga@epfl.ch}
\author[1]{François X. Passelègue}
\author[2]{Nicolas Brantut}
\author[3]{Fabian Barras}
\author[1]{Mathias Lebihain}
\author[1]{Marie Violay}

\address[1]{Laboratoire de M\'ecanique des Roches, \'Ecole Polytechnique F\'ed\'erale de Lausanne, Switzerland}
\address[2]{Department of Earth Sciences, University College London, London, UK}
\address[3]{NJORD Centre for Studies of the Physics of the Earth, University of Oslo, Norway}

\begin{abstract}
Potential energy stored during the inter-seismic period by tectonic loading around faults is released during earthquakes as radiated energy, heat and fracture energy. The latter is of first importance since it controls the nucleation, propagation and arrest of the seismic rupture. On one side, the fracture energy estimated for natural earthquakes (also called breakdown work) ranges between 1 $\mathrm{J/m^2}$ and tens of $ \mathrm{MJ/m^2} $ for the largest events, and shows a clear slip dependence. On the other side, recent experimental studies highlighted that, at the scale of the laboratory, fracture energy is a material property (energy required to break the fault interface) limited by an upper bound value corresponding to the fracture energy of the intact material (1 to 10 $ \mathrm{kJ/m^2} $) independently of the size of the event, i.e. of the seismic slip. 
To reconcile these contradictory observations, we performed stick-slip experiments, as analog for earthquakes, in a bi-axial shear configuration. We analyzed the fault weakening during frictional rupture by accessing to the near-fault (1 mm away) stress-slip curve through strain-gauge array. We first estimated fracture energy by comparing the experimental strain with the theoretical predictions from both Linear Elastic Fracture Mechanics (LEFM) and a Cohesive Zone Model (CZM). By comparing these values to the breakdown work obtained from the integration of the stress-slip curve, we show that, at the scale of our experiments, fault weakening is divided into two stages; the first one, corresponding to an energy of few $ \mathrm{J/m^2} $, consistent with the estimated fracture energy (by LEFM and CZM), and a long-tailed weakening corresponding to a larger energy not localized at the rupture tip, increasing with slip.
Through numerical simulations, we demonstrate that only the first weakening stage controls the rupture initiation and that the breakdown work induced by the long-tailed weakening can enhance slip during rupture propagation and allow the rupture to overcome stress heterogeneity along the fault. We conclude that the origin of the seismological estimates of breakdown work could be related to the energy dissipated in the long-tailed weakening rather than to the one dissipated near the tip.
\end{abstract}


\end{frontmatter}


\section{Introduction}
Earthquakes are due to the abrupt release of part of the elastic stored energy accumulated during the inter-seismic period, which is released as radiated energy in the bulk and dissipated energy in the vicinity of the fault. The latter can be subdivided into two contributions: (1) the so-called breakdown work, which is associated to fault weakening down to some minimum frictional strength, and (2) the remaining frictional dissipation \citep{Kanamori1977,Kanamori2004}. The breakdown work is a collective dissipation term that includes on- and off-fault processes occurring at a range of timescales during rupture, from the onset (i.e., near the tip of the propagating rupture) to the later stages of slip (i.e., far from the tip). Inspired from the energy budget of slip-weakening models of earthquakes \citep[e.g.,][]{palmer1973growth}, the breakdown work is often proposed as a proxy for the fracture energy \citep{Venkataraman2004,Abercrombie2005}, which is defined as the energy consumed at the rupture tip to propagate the rupture by a unit area. However, breakdown work is identified to fracture energy only if fault weakening is concentrated near the propagating edge of the rupture, which is not expected to be systematically the case during natural earthquakes \citep[e.g.,][]{Lambert2020,brener2020unconventional}. How this dissipated energy is distributed around the propagating rupture has a key impact on its dynamics.

Estimating the partitioning and spatio-temporal distribution of energy dissipation during earthquakes is of first importance since they control the nucleation and propagation of the seismic rupture, as well as the intensity of the wave radiation at the origin of ground motions. Unfortunately, seismological observations do not allow for a complete estimate of the energy balance of crustal earthquakes, due to the presence of several unknowns, such as the stresses acting on the fault and the local seismic slip. The analysis of the radiated seismic waves provides a good estimate of the radiated energy \citep{Kanamori1977,Venkataraman2004}, but quantifying the breakdown work of earthquakes remains challenging and relies on a number of simplifying assumptions that are difficult to assess. Seismological estimates indicate that breakdown work scales with earthquake slip, as power law with exponent ranging from 0.5 to 2 \citep[e.g.,][]{Abercrombie2005, Viesca2015}. 

Laboratory studies have brought useful constraints on the energetics of shear ruptures \citep[e.g.,][]{Johnson1976,ohnaka1989cohesive,rubinstein2004detachment,Svetlizky2014,Bayart2016,Xu2019}. Stick-slip experiments conducted in rocks or other materials have shown that the onset of frictional slip can be described by a shear crack (i.e., mode II fracture) nucleating and propagating along the fault interface. Using Linear Elastic Fracture Mechanics (LEFM), recent studies \citep[e.g.,][]{Svetlizky2014, Bayart2016, kammer2019fracture} have highlighted that the stress field and associated release of elastic energy at the rupture tip is fully controlled by an effective fracture energy of the interface that is a scale-independent material property. Laboratory-derived estimates yield upper bound values for the shear fracture energy that are commensurate to the mode I fracture energy of the intact material (1 to 10 $ \mathrm{kJ/m^2} $). Remarkably, when frictional motions can be described by fracture mechanics, the nucleation of the instability can be modeled by rate and state frictional laws, assuming common values for the rate and state parameters, that respect the shear fracture energy estimated at the rupture tip during its propagation \citep{Latour2013,Kaneko2016}.
In addition, the propagation and arrest of dynamic ruptures in laboratory samples has been shown to be fully described by fracture mechanics  \citep{Kammer2015,Bayart2016,Galis2017,Svetlizky2014,Passelegue2020}, raising the hope of predicting earthquake motions.

However, laboratory studies have shown values of fracture energy of the order of tenths to hundreds of kJ/m$^2$, far from those of natural earthquakes ($ \mathrm{MJ/m^2} $), suggesting a dichotomy between the processes occurring at the two scales of observations.  
Effectively, at the scale of natural faults, seismological observations indicate a slip-dependence of the breakdown work of earthquakes \citep{Abercrombie2005}, with values ranging from 1 $ \mathrm{J/m^2} $ to tens of $ \mathrm{MJ/m^2} $ for the largest crustal earthquakes (i.e. three to four order of magnitude larger than the fracture energy of intact material constituting the seismogenic crust), differing from the notion of fracture energy as a constant material property. 
Recent work by \cite{Ke2020} suggests that apparent scale-dependent breakdown work can emerge in ruptures governed by an underlying constant (material-dependent) fracture energy when earthquakes propagate into regions of decreasing background stress, where ruptures progressively stop. Such apparent scaling arises due to stress drop heterogeneity rather than intrinsic fault strength evolution. %

By contrast with laboratory \emph{rupture} experiments, \emph{friction} experiments at high slip velocity, aimed at characterizing the evolution of frictional stress that would be observed at a single point along the fault during seismic slip and have reproduced the slip-dependence of breakdown work \citep{nielsen2016g,cornelio2020effect,passelegue2016dynamic}. Similarly, fault models based on weakening mechanisms such as thermal pressurization \citep{Viesca2015, Lambert2020} or flash heating \citep{Brantut2017} have also been shown to exhibit scaling between slip and breakdown work. In both experiments and models, most of the total dissipated energy is converted into heat, further enhancing the weakening of the fault during coseismic slip due to the occurrence of thermally activated weakening processes. In this regard \cite{Lambert2020} emphasize how, due to this enhanced fault weakening prolonged after rupture propagation, breakdown work does not solely correspond to dissipation occurring within a small region near the propagating rupture edge ( cohesive zone), but includes possibly large contributions from dissipation occurring at large distances from it. The exact role of such ``long-tailed'' weakening in the dynamics of rupture propagation, and in particular its possible contribution to an effective fracture energy at the propagating tip, remains somewhat unclear. Using rate-and-state models of friction, recent works show that while the dynamics of the frictional rupture can be described by fracture mechanics, the fracture energy inverted at the crack edge only corresponds to a smaller fraction of the breakdown work integrated during rupture \citep{barras2020emergence,brener2020unconventional}. \\
In this paper, we combine, in a single experimental setup, the study of rupture dynamics and friction evolution. From the variation of frictional stress with slip measured in the wake of the rupture, we show that the fracture energy only represents a small fraction of the total breakdown work at the scale of laboratory experiments. Building on these observations, this manuscript tackles two objectives: firstly, to investigate and quantify the discrepancy between fracture energy and breakdown work existing at the scale of laboratory experiments, and secondly to discuss how the observed dynamics can be up-scaled to the energy budget of earthquakes.

\section{Methods}
\subsection{Apparatus and loading conditions}
Experiments were performed with a bi-axial shear apparatus, located at LEMR EPFL. The apparatus is composed of a rigid steel frame holding two rectangular cuboid blocks of polymethylmethacrylate (PMMA) of known elastic properties (Young's modulus $E$=5.7 GPa and Poisson's ratio $\nu$=0.33) (Figure\ref{fig:1}a.). The dimensions of the PMMA blocks are of 20~cm $\times$ 10~cm $\times$ 1~cm for the upper block, and 50~cm $\times$ 10~cm $\times$ 3~cm for the lower block, resulting in a 20~cm $\times$ 1~cm fault interface. External loading is imposed by using two Enerpac handpumps applying respectively normal and shear load with a maximum stress of 20~MPa (Figure \ref{fig:1}a). The applied macroscopic loads were measured using two load cells located between the frame and the pistons, and recorded at 100~Hz sampling rate with a National Instrument data acquisition system.
To capture the details of the dynamic ruptures, the upper PMMA block was equipped with an equally spaced array of strain gauge rosettes placed 1~mm away from the fault, which guaranteed high frequency measurement of strain (for details on the acquisition system refer to the Supplementary Material).
\subsection{Experimental protocol}
To reproduce earthquakes with our experimental system, a normal load was first imposed along the fault, up to values ranging between 0.2 and 5 MPa. Then, the shear load was manually increased up to the onset of instability, which resulted in a fast release of stress along the experimental fault, associated with seismic slip and elastic wave radiation (i.e., stick slip events).

\subsection{Estimation of local strain and rupture velocity}
\begin{figure}
  \includegraphics[width=\linewidth]{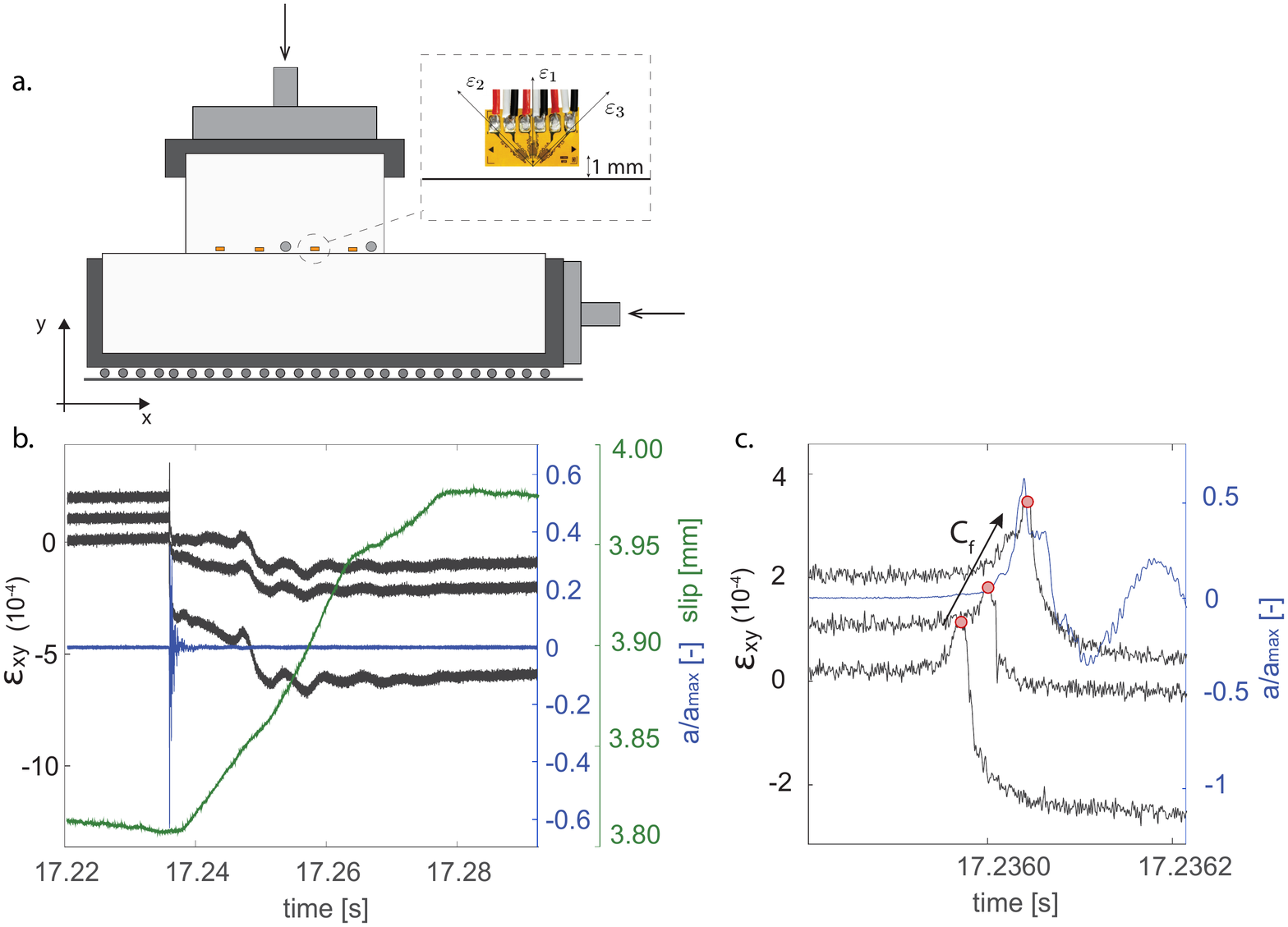}
  \caption{a. Sketch of the biaxial apparatus used to perform stick-slip experiments. Symbols as rectangles and circles represent respectively strain gauges and accelerometers placed at a distance of 1 mm from the fault. b. Evolution of strain (in black) during the occurrence of a rupture event. Strain is measured through strain gauge rosettes placed at three different location along the fault. In green the macroscopic slip evolution measured through laser displacement sensor is shown. Macroscopic slip is initiated once rupture has propagated all the way through the fault. The acceleration evolution (in blue) shows radiation occurring mainly during rupture propagation and dissipating as macroscopic slip occurs. c. Zoom of strain and acceleration distributions during the rupture event. Rupture arrival times for each strain rosette (in red) used to estimate the rupture velocity.}
  \label{fig:1}
\end{figure}

During stick-slip instabilities, the local material response was analyzed using the strain gauge array. Denoting $x$ and $y$ the fault-parallel and the fault-perpendicular coordinates, respectively, the elements $\varepsilon_\mathrm{xx}, \varepsilon_\mathrm{yy}, \varepsilon_\mathrm{xy}$ of the strain tensor were obtained from the measured strain (referred to as $\varepsilon_\mathrm{1}, \varepsilon_\mathrm{2}, \varepsilon_\mathrm{3}$  for strain gauges oriented at $ 45^{\circ}$, $90^{\circ}$ and $135^{\circ} $ from the fault direction, respectively) as
\begin{equation}
\begin{aligned}
	 \varepsilon_\mathrm{yy} &=\varepsilon_\mathrm{1}, \\
	 \varepsilon_\mathrm{xy} &=\frac{\varepsilon_\mathrm{3}-\varepsilon_\mathrm{2}}{2}, \\
	 \varepsilon_\mathrm{xx} &=\varepsilon_\mathrm{3}+\varepsilon_\mathrm{2}-\varepsilon_\mathrm{1}.
\end{aligned}
\end{equation}
Typical time series of shear strain ($\varepsilon_\mathrm{xy}$) computed at each rosette location, together with the laser displacement sensor and the acceleration motions, during a stick-slip instability (here obtained at 2.3 MPa normal stress) are presented in Figure \ref{fig:1}b.

Rupture velocity ($C_\mathrm{f}$) was estimated using the times at which the passage of rupture front was detected from the different strain gauges and the relative distance between them. The arrival of the rupture front was determined as the moment at which the strain gauges signal reached its peak (Figure \ref{fig:1}c); this method assumes that the rupture speed is constant over the distance spanned by the gauge array. An increase in rupture velocity is observed with an increase in the initial peak shear stress, as observed in previous studies \citep{ben2010dynamics,passelegue2016dynamic}.
Once the rupture fully propagated along the interface, the two sides of the fault started behaving like rigid blocks slipping one against the other, as shown by the evolution of the macroscopic slip and the cessation of measured acceleration motions (Figure \ref{fig:1}b).

\section{Experimental Results}

\subsection{Estimation of the fracture energy}
\begin{figure}
\centerline{\includegraphics[width=\linewidth]{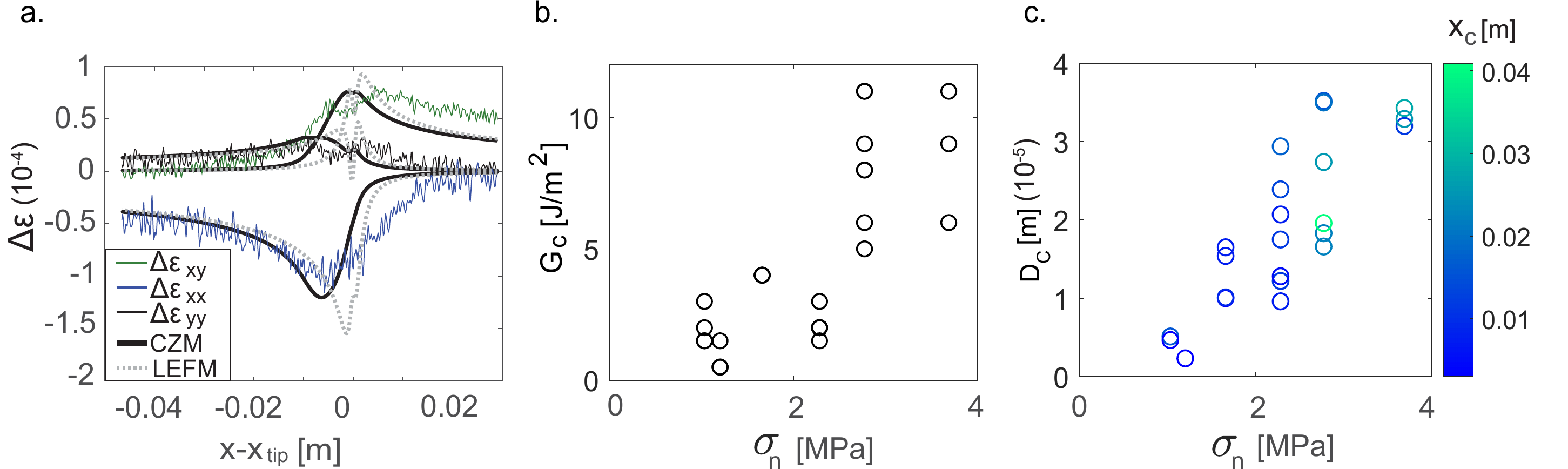}}
  \caption{a. Strain variation evolution during a rupture event ($\Delta\varepsilon_\mathrm{xx}, \Delta\varepsilon_\mathrm{yy}, \Delta\varepsilon_\mathrm{xy}$). Theoretical predictions from CZM (in black) and LEFM (dashed gray) are plotted as well. b. Evolution of fracture energy inverted for different events for increasing applied normal load. c. Critical distance ($D_\mathrm{c}$) evolution with applied normal load, obtained by making use of cohesive zone ($ x_\mathrm{c} $) inverted through CZM.}
  \label{fig:2}
\end{figure}

Using the measured rupture velocities, dynamic strain perturbations recorded at the passage of the rupture tip were compared to theoretical predictions using both a Cohesive Zone Model (CZM) \citep{Poliakov2002a,kammer2019fracture} and LEFM \citep{Svetlizky2014} (Supplementary material).
The LEFM solution was fitted by adjusting a single parameter, the stress intensity factor $K_\mathrm{II}$, while the CZM solution was fitted by adjusting the stress drop $(\tau_\mathrm{p}-\tau_\mathrm{r})$ with $\tau_\mathrm{p}$ the peak strength and $\tau_\mathrm{r}$ the residual stress and the cohesive zone length $x_\mathrm{c}$ (Figure \ref{fig:2}a). Both LEFM and CZM predictions output identical values of $G_\mathrm{c}$. This inversion was done for several events occurring at different applied normal loads (Figure \ref{fig:2}b). As expected by previous studies  \citep{okubo1984effects,Bayart2016}, $G_\mathrm{c}$ increases with increasing applied normal load, due to an increase of contact area between the two surfaces. The values found ranged between 0.5 and 11 $ \mathrm{J/m^2} $, in agreement with previous estimates \citep{Svetlizky2014,Bayart2016}.
Our results suggest that the cohesive zone (inverted from CZM) increases with the initial normal stress applied, with values ranging from 1 to 10 mm at 0.2 and 4 MPa applied normal stress, respectively (Figure \ref{fig:2}c). Note that for events presenting small values of $x_\mathrm{c}$, CZM predictions collapse to those of LEFM, as expected theoretically and previously observed \citep{Svetlizky2014}. Finally, using our experimental estimates of the rupture velocity and our theoretical predictions for $x_\mathrm{c}$, a characteristic slip weakening distance was estimated as \citep{palmer1973growth}
\begin{equation}
\label{eqn:dc}
	 	 D_\mathrm{c}= x_\mathrm{c}\, 4\,(1-\nu) \,(\tau_\mathrm{p}-\tau_\mathrm{r})\,/\,\pi\mu.
\end{equation}
$D_\mathrm{c}$ increases with the initial normal stress from a few microns at the lowest stress tested to tens of microns at $\approx 4$ MPa normal stress (Figure \ref{fig:2}c.), in agreement with previous studies \citep{ohnaka2003constitutive,passelegue2016dynamic}.

\subsection{Comparison to local slip measurements}
The values of fracture energy and frictional parameters inverted from CZM can be compared to the local evolution of stress versus slip. First of all, using the local strain tensor and the material's elastic properties, under the assumption of plain strain conditions, the shear stress evolution ($\tau$) during instability was computed at 1~mm from the fault. Secondly, the strain measurements were used to compute the local slip induced along the fault during rupture propagation. The particle velocity was estimated from the strain component parallel to the slip direction, following
$ \dot{u}_\mathrm{x}=- C_\mathrm{f}\varepsilon_\mathrm{xx} $ \citep{Svetlizky2014}. Then, local fault slip was obtained by integrating $\dot{u}_\mathrm{x}$ with respect to time. The latter was compared to the slip obtained from the calibrated accelerometers located along the fault, computed following  $ u_\mathrm{x}= \iint_\mathrm{t} a(x) \,dt $, with $ a $ the measured acceleration in m/s$^{2}$ and $ t$ the time during propagation. The evolution of slip during rupture propagation obtained from both strain gauges and accelerometers is comparable. The final values of slip obtained in this way are also comparable to the macroscopic slip measured by the laser sensor, suggesting that strain gauges provide a robust estimate of the local slip during rupture propagation, excluding possible strain-induced waves reflection. The total displacement occurring on the fault was computed as $ D(t)=2 u_\mathrm{x} $, assuming a symmetric displacement across the fault, given the uniform far-field loading. 

\begin{figure}
  \includegraphics[width=\linewidth]{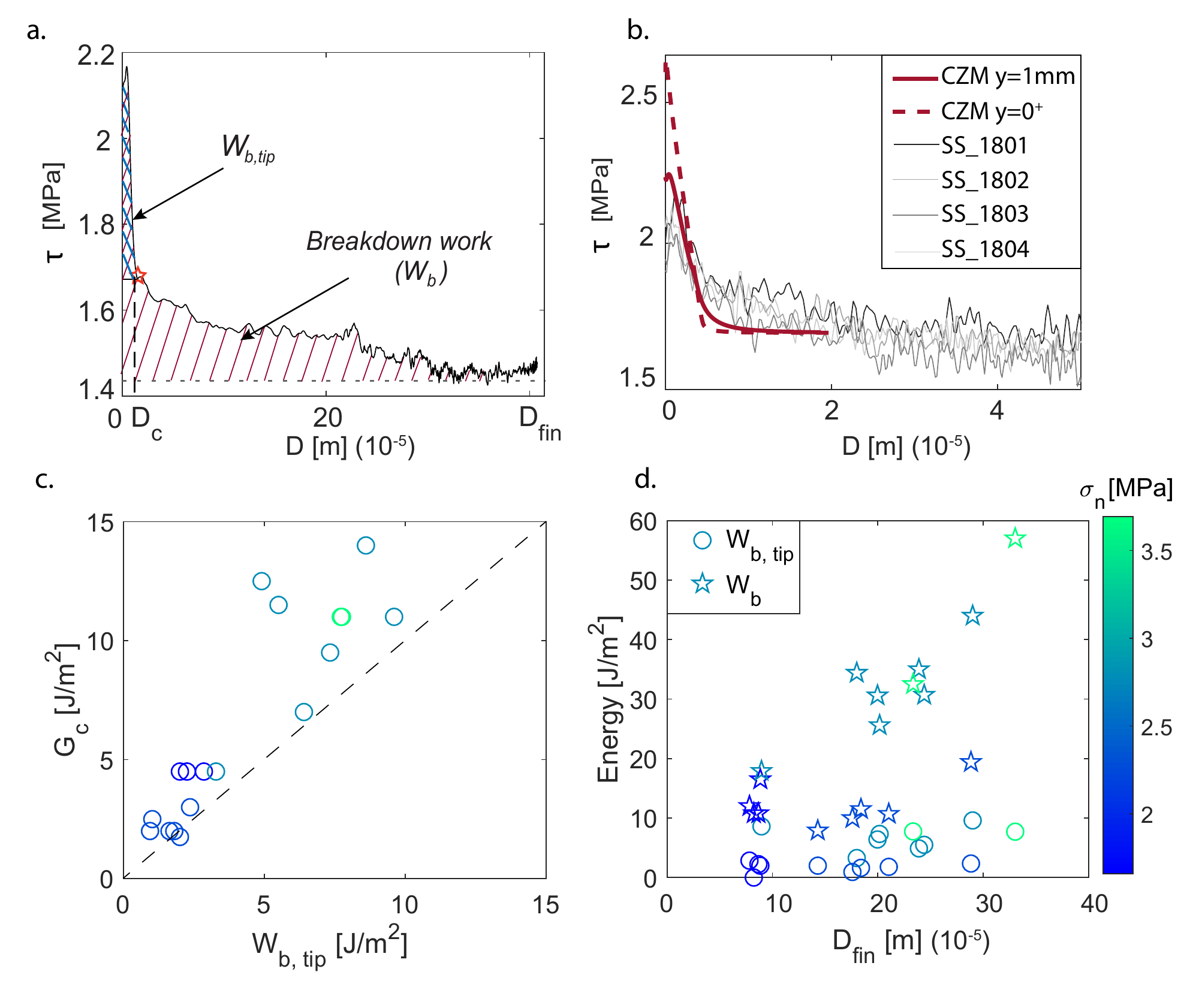}
  \caption{a. Evolution of shear stress with fault slip for a specific rupture event. The area in blue represents the near tip breakdown work ($ W_\mathrm{b, tip} $), the one in red the overall breakdown work ($ W_\mathrm{b} $). b. Theoretical predictions from CZM corresponding to a distance of 1 mm from the fault (solid red line) and of $ 0^+$ mm (dashed red line) plotted with the experimental curves. c. Comparison between fracture energy obtained from theoretical inversions $ G_\mathrm{c}$ Figure \ref{fig:2} and $ W_\mathrm{b, tip} $. d. Evolution of $ W_\mathrm{b, tip} $ and  $ W_\mathrm{b}$ with applied normal load and associated final slip.}
  \label{fig:3}
\end{figure}

In agreement with the slip weakening assumption used in CZM, the onset of slip is marked by a large stress release (around 0.5 MPa) within a small amount of slip (around $10$~$\mu$m) (Figure \ref{fig:3}a), an outcome which is in good agreement with our estimates of $D_\mathrm{c}$ using equation \eqref{eqn:dc}. This abrupt weakening stage is followed by a second long-tailed weakening stage during which the stress decreases continuously  with increasing slip, at a much lower rate. During the first weakening stage, $70\% $ of the final stress drop is achieved in the first micrometers of slip (Figure \ref{fig:3}). During the second stage, the weakening  continues in a less severe manner until the arrest of dynamic slip, defined here as the time at which the rupture propagated through the entire fault. While the first weakening stage is predicted by CZM at the strain gauge locations (Figure \ref{fig:3}b), this long-tailed weakening is not expected to occur from the model, suggesting that at the scale of our experiments, fault weakening is more complex than expected from linear slip-weakening model. This dual-scale weakening has been observed for decades in studies of engineering materials like concrete \citep{Planas1997,Bazant2004}, and is expected to give rise to a scale-dependent fracture energy, as it is observed from earthquakes scaling law \citep{madariaga2009}. 

\section{Discussion}
 \subsection{Comparison between fracture energy, near-edge and long-tailed breakdown work}
Keeping these last observations in mind, we now assume that the evolution of stress and slip estimated using the strain gauges located at 1 mm from the fault are representative of the real motions occurring along the fault during rupture motions. This assumption seems robust since (i) the slip inverted from strain gauges at 1 mm from the fault is comparable to the one measured by the accelerometers and the laser sensor, (ii) the evolution of the stress 1 mm away from the fault is close to the evolution of the stress on the fault, particularly in terms of energy dissipated (Figure \ref{fig:3}b). In general, off-fault shear stress is similar to that on the fault when it is measured at distances much smaller than the size of the cohesive zone, which is verified here.

The estimates of $ D_\mathrm{c} $ allowed us to differentiate two principal weakening stages and to compute the energy dissipated during each of them. The energy dissipated at the crack edge, also known as the edge-localized dissipation \citep{barras2020emergence}, was computed for each event as  $W_\mathrm{b,tip} = \int_0^{D_\mathrm{c}} \left(\tau\left(D\right) - \tau\left(D_\mathrm{c}\right)\right) dD$, using the measured shear stress $\tau$. These values are in good agreement with  $ G_\mathrm{c} $ estimates obtained from the direct inversion of strain variation measurements shown above (Figure \ref{fig:3}c), showing that our near-fault stress measurements can be considered representative of on-fault stress, and, once more, that dynamic fracture mechanics is able to describe the onset of frictional sliding.

Secondly, the total dissipated energy resulting from the full stress evolution (i.e. breakdown work) was computed following
\begin{equation}
\label{eqn:WB}
		 W_\mathrm{b} = \int_0^{D_\mathrm{fin}} \left(\tau\left(D\right) - \tau\left(D_\mathrm{fin}\right)\right) dD
\end{equation}
where $D_\mathrm{fin}$ corresponds to the final value of slip recorded during rupture propagation. The energy dissipated during the complete weakening processes presents final values ranging between 1 and 60 $\mathrm{J/m^{2}}$, i.e. values that are one order magnitude greater than $W_\mathrm{b,tip}$ and $G_\mathrm{c}$. While $G_\mathrm{c}$ slightly increases with applied normal load, as already discussed, $W_\mathrm{b}$ covers a much wider range of values, systematically higher than $G_\mathrm{c}$, and presents a clear dependence with the final slip (Figure \ref{fig:3}d). These observations suggest that contrary to the energy dissipated at the rupture edge, which can be considered as a normal contact problem (fault roughness, normal pressure), the energy dissipated during the second weakening stage is rather controlled by frictional processes and slip history, presenting features similar to the breakdown work derived from high velocity friction experiments \citep{nielsen2016g} and natural earthquakes \citep{Abercrombie2005}. At the scale of our experiments, this large breakdown work does not contribute to the propagation and the dynamics of the rupture tip, since the stress intensity recorded by the strain gauges is controlled by $W_\mathrm{b,tip}$. However, this could be related to our finite fault length which is small compared to an effective cohesive zone related to the long-tailed weakening stage.

\subsection{Theoretical implications of long-tailed breakdown work on rupture dynamics}
In our experiments, the prolonged weakening does not contribute to fracture energy. However, one may wonder how and at which scale the long-tailed weakening may control rupture dynamics. As a first step, we analyze theoretically the influence of the cohesive stress distribution on the stress intensity factor, and examine how stress variations far from the rupture tip may actually contribute to tip dynamics.

Let us consider a semi-infinite straight crack nucleating at t=0 in an infinite elastic medium. The crack is loaded under anti-plane shear conditions with a constant uniform background stress $\tau_\mathrm{b}$. The propagation of the shear crack is resisted by cohesive frictional stresses $\tau_\mathrm{f}\left(x,t\right)$. Following our experimental results, which provide evidence for a dual-scale weakening stage, the frictional stresses can be decomposed into the sum of three terms defined by (i) $\tau_\mathrm{f,tip}\left(D(x,t)\right)$ describing the near-tip weakening due to the local instantaneous slip $D$, (ii) $\tau_\mathrm{f,tail}\left(D(x,t)\right)$ associated to the long-tailed weakening, and (iii) the uniform residual stresses $\tau_\mathrm{r,tail}$ at large slip. The stress intensity factor resulting of the evolution of stress with slip is as \citep{Kostrov1966}

\begin{flalign}
\label{eqn:Ktot}
k_\mathrm{tot}(x_\mathrm{tip}, C_\mathrm{f},t) & = \beta_\mathrm{s}\left(C_\mathrm{f}\right) \int_{0}^{C_\mathrm{s}t} \left[\tau_\mathrm{b}\left(x_\mathrm{tip}-r\right)-\tau_\mathrm{r,tail}\right] \frac{dr}{\sqrt{r}} &\nonumber\\
& - \beta_\mathrm{s}\left(C_\mathrm{f}\right)\int_{0}^{C_\mathrm{s}t} \tau_\mathrm{f,tip}\left(D(x_\mathrm{tip}-r,t-r/C_\mathrm{s})\right)   \frac{dr}{\sqrt{r}} &\\
& - \beta_\mathrm{s}\left(C_\mathrm{f}\right) \int_{0}^{C_\mathrm{s}t} \tau_\mathrm{f,tail}\left(D(x_\mathrm{tip}-r,t-r/C_\mathrm{s})\right)  \frac{dr}{\sqrt{r}}. &\nonumber\\
\end{flalign}

where $x_\mathrm{tip}$ is the position of the rupture tip, $\beta_\mathrm{s}\left(C_\mathrm{f}\right)=\sqrt{\frac{2}{\pi}}\sqrt{1-C_\mathrm{f}/C_\mathrm{s}}$ is a universal pre-factor depending of the crack velocity $C_\mathrm{f}$, $r$ is the longitudinal distance to the crack tip, and $C_\mathrm{s} $ is the shear wave speed of the material. 

The presence of cohesion behind the rupture tip implies that the stresses remain non-singular at the crack tip ($k_\mathrm{tot}=0$). Assuming this, the total stress intensity factor can then be rewritten from equation \eqref{eqn:Ktot} as 
\begin{equation}
\label{eqn:Kcontributions}
k_\mathrm{tot}(x_\mathrm{tip}, C_\mathrm{f},t) = k(x_\mathrm{tip},C_\mathrm{f}) - k_\mathrm{tip} (x_\mathrm{tip},C_\mathrm{f},t) - k_\mathrm{tail}(x_\mathrm{tip},C_\mathrm{f},t) = 0,
\end{equation}
where $k$ is the stress intensity factor that emerges when all weakenings are occurring within an infinitesimally small region behind the crack tip,  $k_\mathrm{tip} $ is the contribution of the near-tip weakening frictional stresses, and $k_\mathrm{tail}$ that of the frictional stresses that weaken far from the rupture tip.

The different terms of equation \eqref{eqn:Kcontributions} are very different by nature. Indeed, since $\tau_\mathrm{f,tip}$ is nonzero only in a small region of dimension $x_\mathrm{c,tip}$ near the tip, $k_\mathrm{tip}$ is independent from time and can be written as a speed-dependent ``cohesion modulus'' $k_\mathrm{tip}(C_\mathrm{f})$ (i.e. dynamic toughness) \citep{Kostrov1966}. On the contrary, the contribution $k_\mathrm{tail}$ of the long-tailed weakening to the total stress intensity factor  $k_\mathrm{tot}$ relates to the distribution of frictional stress $\tau_\mathrm{f,tail}$ in a larger region of size $x_{c,tail} \gg x_{c,tip}$ with some delay due to the wave-mediated nature of the stress redistribution. As such, if the breakdown work $W_\mathrm{b}$ of equation \eqref{eqn:WB} depends only on the total slip $D_\mathrm{fin}$, the energy dissipated at the tip depends on the spatio-temporal evolution of slip $D\left(x,t\right)$ during rupture propagation. One may then distinguish two characteristic regimes where (i) only the first weakening $\tau_\mathrm{f,tip}$ participates in the rupture dynamics and the rupture energy balance, and (ii) the two weakenings $\tau_\mathrm{f,tip}$ and $\tau_\mathrm{f,tail}$ are both involved.

When the crack and the amount of slip $D$ are small, or when the crack velocity $C_\mathrm{f}$ is large, the long-tailed weakening is not activated or its information in the crack wake does not have time to get to the crack tip, $\tau_\mathrm{f,tail}$ of equation \eqref{eqn:Ktot} remains approximately constant outside of the first cohesive zone. Following equation \eqref{eqn:Kcontributions}, this large-scale weakening is not perceived by the propagating crack, and does not feed its dynamics. Consequently, the stress singularity in front of the the crack tip is dominated by $\propto k_\textsc{lefm}/\sqrt{r}$, where $k_\textsc{lefm} = (k - k_\mathrm{tail}) = k_\mathrm{tip}(C_\mathrm{f})$. Assuming this hypothesis, the dynamic energy balance can be written following (\cite{Freund1998} chap. 5): 
\begin{equation}
\label{eqn:tipdissipation}
G = \frac{k_\textsc{lefm}(x_\mathrm{f},C_\mathrm{f})^2}{2\mu\sqrt{(1-C_\mathrm{f}^2/C_\mathrm{s}^2)}} = \frac{k_\mathrm{tip}(C_\mathrm{f})^2}{2\mu\sqrt{(1-C_\mathrm{f}^2/C_\mathrm{s}^2)}} = G_\mathrm{c,tip},
\end{equation}
meaning the energy dissipated to make the crack propagate corresponds to the near-tip fracture energy only. Note that once the long tailed weakening initiates, the stress state in the vicinity of the crack tip is excepted to results from the combination of background stress and long tailed frictional stress following $\tau \propto k-k_\mathrm{tail}$. However in
this intermediate case, no clear residual frictional stress is achieved during propagation, so that the energy balance at the rupture tip is only approximately described by equation \eqref{eqn:Kcontributions}. \\
On the contrary, when the crack is long enough so that both the near-tip and long-tailed weakening occur within a small region behind the crack tip, a well defined residual stress $\tau_\mathrm{r,tail}$ is reached behind the crack, and this information can reach the crack tip. In that case, both types of weakening feed the rupture dynamics, and the stress singularity is given by $k_\textsc{lefm} = k$. The energy balance reads
\begin{equation}
\label{eqn:wakedissipation}
G = \frac{k_\textsc{lefm}(C_\mathrm{f})^2}{2\mu\sqrt{(1-C_\mathrm{f}^2/C_\mathrm{s}^2)}} = G_\mathrm{c,tail}+G_\mathrm{c,tip}.
\end{equation}
In that case, the fracture energy measured from the tip stresses now equals to the complete breakdown work, potentially much larger than the fracture energy associated with the near-tip weakening.
Overall, our analysis shows that the impact of the prolonged weakening at large slip (and large distances from the crack tip) on the rupture dynamics is wave-mediated, and as such, may not be significant enough to impact the crack tip energy balance. For short crack lengths or near-$C_\mathrm{s}$ rupture, crack dynamics are dominated by the near-tip weakening only, and the total breakdown work can be much larger than fracture energy. For large rupture lengths and slow ruptures, we return to a classical situation where breakdown work and fracture energy are equal. The transition between these two simple regimes is quantitatively described in the next section.

\subsection{Contributions of long-tailed weakening in the energy release rate}
\begin{figure}
\centerline{\includegraphics[width=\linewidth]{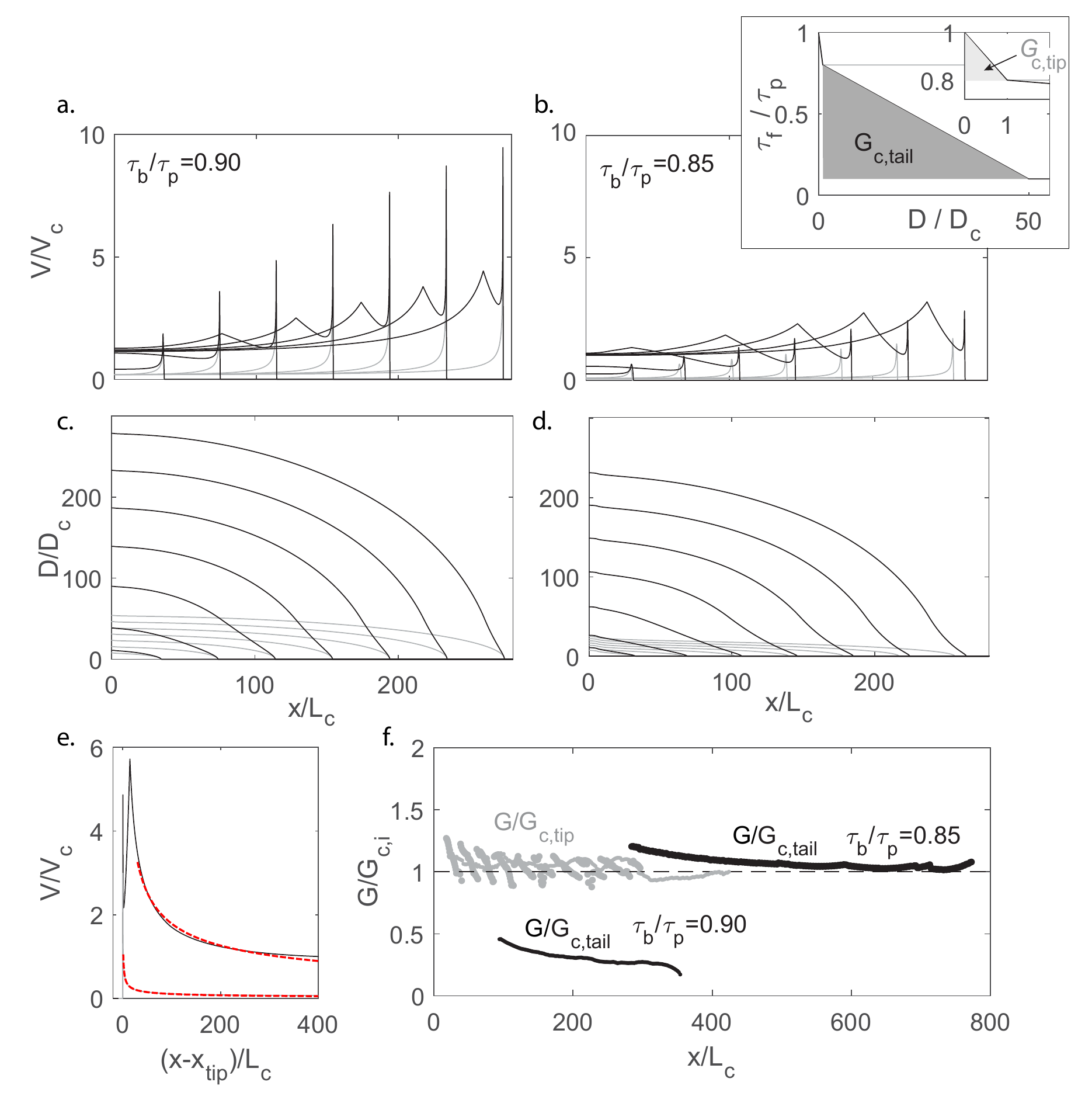}}
  \caption{a., b. Normalized slip rate ($ V/V_\mathrm{c} $) evolution along fault length for simple (in gray) and dual-scale (in black) slip weakening laws, respectively for scenario S1 ($\tau_\mathrm{b}/\tau_\mathrm{p}=0.90$) and S2 ($\tau_\mathrm{b}/\tau_\mathrm{p}=0.85$). $V_\mathrm{c}=\mu C_s /\tau_\mathrm{p}$ is the critical slip rate. Inset: constitutive laws used for the numerical simulations. In gray the simple slip weakening law describing the first weakening stage observed, in black the dual-scale slip weakening law describing both first and second weakening with associated fracture energies ($ G_\mathrm{c,tip}, G_\mathrm{c,tail} $). c.,d. Slip profile evolution along fault length for both weakening laws for scenario S1 and S2. e. Example of fit of slip rate profiles with theoretical predictions (in dashed red) for the simple weakening case and dual-weakening case with $\tau_\mathrm{b}/\tau_\mathrm{p}=0.85$. f. Energy release rate evolution with rupture size for the simple weakening law normalized by fracture energy $ G_\mathrm{c,tip} $ (in gray) and for the dual-scale weakening law normalized by fracture energy $ G_\mathrm{c,tail} $ (in black) for scenario S1 and S2.}
  \label{fig:4}
\end{figure}

Once the rupture length ($L_\mathrm{f} $) reaches a sufficient size $L_\mathrm{f}\gg x_\mathrm{c,tail}$, two scenarios are admissible in light of the small-scale yielding requirement (i.e. dissipative phenomena limited to a region much smaller than the dimensions of the system). In a first one (named hereafter S1), the dissipation length characterizing the first weakening mechanism is much smaller than the one of the second weakening $x_\mathrm{c,tip}\ll x_\mathrm{c,tail}$. A rupture driven exclusively by the first weakening stage could still experience edge-localized dissipation and propagate ahead and almost independently of the long-tailed weakening mechanism (similar to what is observed with rate-dependent friction \citep{brener2020unconventional}). On the other hand (scenario S2), the first weakening stage becomes a negligible detail inside the large process zone and the macroscopic rupture dynamics is dominated by the larger fracture energy $G_\mathrm{c,tail}$. To shed light on the realization of these two scenarios, we conduct numerical simulations of frictional ruptures (see Supplementary Materials for details on the numerical method) driven by slip weakening friction laws with different weakening length scales. For simplicity, only mode III ruptures were studied in order to avoid rupture propagation velocities larger than the shear wave speed, which would add unnecessary complexity to our results. The reference case consists of a linear slip weakening law defined by a peak stress $\tau_\mathrm{p}$, residual stress $\tau_\mathrm{r} = 0.8\tau_\mathrm{p}$ and a slip-weakening distance $D_\mathrm{c, tip}$. The tested case consists of a dual-scale slip weakening law, that matches the reference case in the first stage, but which is followed by a second long-tailed weakening stage (Figure \ref{fig:4} inset) allowing a larger stress release up to a final residual stress $\tau_\mathrm{r,tail} = 0.1 \tau_\mathrm{p} $ over a weakening distance $D_\mathrm{c,tail }=50 D_\mathrm{c, tip} $. In both cases, the initial background stress ($\tau_\mathrm{b}$) along the fault was set to a uniform value, and rupture nucleation was triggered by imposing an elevated stress patch $\tau_\mathrm{b,nucl}$ 5\% above $\tau_\mathrm{p}$ in a small region at the center of the modeled fault. \\
During the propagation phase of the rupture, the numerical results obtained for the reference slip weakening law show a symmetric crack-like rupture propagating across the interface, with an increase in stress and slip velocity occurring near the edge of the crack (Figures \ref{fig:4}a and \ref{fig:4}c). To further investigate the dynamics at the rupture tip, the increase in slip velocity at the vicinity of the crack edge was fitted with LEFM predictions (Figure \ref{fig:4}d) following \citep{barras2020emergence}
\begin{equation}
\label{eqn:Vlefm}
  v(r=x-x_\mathrm{tip},\theta=\pi,C_\mathrm{f}) \approx\frac{K_\mathrm{III}^2 C_\mathrm{f}}{\sqrt{2\pi(x-x_\mathrm{tip})}\mu\alpha_\mathrm{s}(C_\mathrm{f})}
\end{equation}
where $K_\mathrm{III}$ is the stress intensity factor, $r,\theta$ is a polar coordinate system moving with the rupture edge, and $\alpha_\mathrm{s}(C_\mathrm{f})=\sqrt{1-C^2_\mathrm{f}/C^2_\mathrm{s}}$. The best fit outputs the solution for the stress intensity factor, which is directly related to the energy release rate following 
\begin{equation}
\label{eqn:grate2}
  G=\frac{K^2_\mathrm{III}}{2\mu\alpha_\mathrm{s}(C_\mathrm{f})}.
\end{equation}
The latter is used to study the near-tip energy balance controlling the dynamics of the rupture tip during its propagation \citep{barras2020emergence}. This analysis demonstrates that during the rupture propagation driven by the simple slip weakening law, the energy balance $G=G_\mathrm{c,tip}$ is systematically respected, independently of the rupture length (Figure \ref{fig:4}f). Note that small variations in the energy release rate are observed during the crack propagation, due to the uncertainties on the estimate of the rupture velocity and sharp variations of $1/\alpha_\mathrm{s}\left(C_\mathrm{f}\right)$ near $C_\mathrm{f} \simeq C_\mathrm{s}$. This result confirms that the energy release rate at the crack tip is controlled by the near-tip fault weakening, as expected theoretically \citep{irwin1957analysis,barras2020emergence}.

Interestingly, the results obtained for the dual-scale weakening law show both of the aforementioned scenarios as function of the background stress. The overall effect of the used dual-scale slip weakening law is reflected in a larger slip and slip velocity in the central part of the crack (Figure \ref{fig:4}), which lead to the emergence of a second increase in slip velocity traveling behind the slip velocity peak characterizing the rupture propagation front. Note that such kind of rupture fronts presenting two successive increases in slip velocity have been recently recorded during rupture experiments presenting low rupture velocities, i.e. low initial normal stress, \citep{berman2020dynamics}.
For frictional rupture under high background stress (i.e. $ \tau_\mathrm{b}=0.9\tau_\mathrm{p} $), the nucleated rupture driven by the first-weakening mechanism ($ G=G_\mathrm{c,tip} $) keeps accelerating such that it is barely perturbed by the effect of the long-tailed weakening. An example of such dynamics corresponding to scenario S1 is presented in Figure \ref{fig:4}a and shows a propagation very similar to the equivalent simple slip weakening setup. Moreover, the increase in the slip velocity profile generated by the long-tailed weakening leads to an associated energy release rate much smaller than $G_\mathrm{c,tail}$, confirming that it is not controlling rupture propagation (Figure \ref{fig:4}f). 
Conversely, if the background stress is smaller (i.e. $ \tau_\mathrm{b}=0.85\tau_\mathrm{p} $), the increase of slip rate generated by the second-weakening stage can reach the leading front and accelerate the rupture further. Such situation is shown in Figure \ref{fig:4}c that highlights how the rupture is now propagating faster than in the case of simple slip weakening law. The inverted value of G from the slip velocity profile is now balancing $  G_\mathrm{c,tail}$, confirming that the long-tailed weakening mechanism is driving the rupture, in agreement with scenario S2. Remarkably, for the slip-weakening model used in these simulations, dynamic fracture arguments can be used to predict the critical level of background stress $\tau^*_\mathrm{b}$ that controls the observed transition between the scenario S1 ($\tau_\mathrm{b}>\tau^*_\mathrm{b}$) and S2 ($\tau_\mathrm{b}<tau^*_\mathrm{b}$)  (see the details in Supplemental material).

\subsection{Contributions of long-tailed weakening in presence of a stress heterogeneity}
\begin{figure}
\centerline{\includegraphics[width=\linewidth]{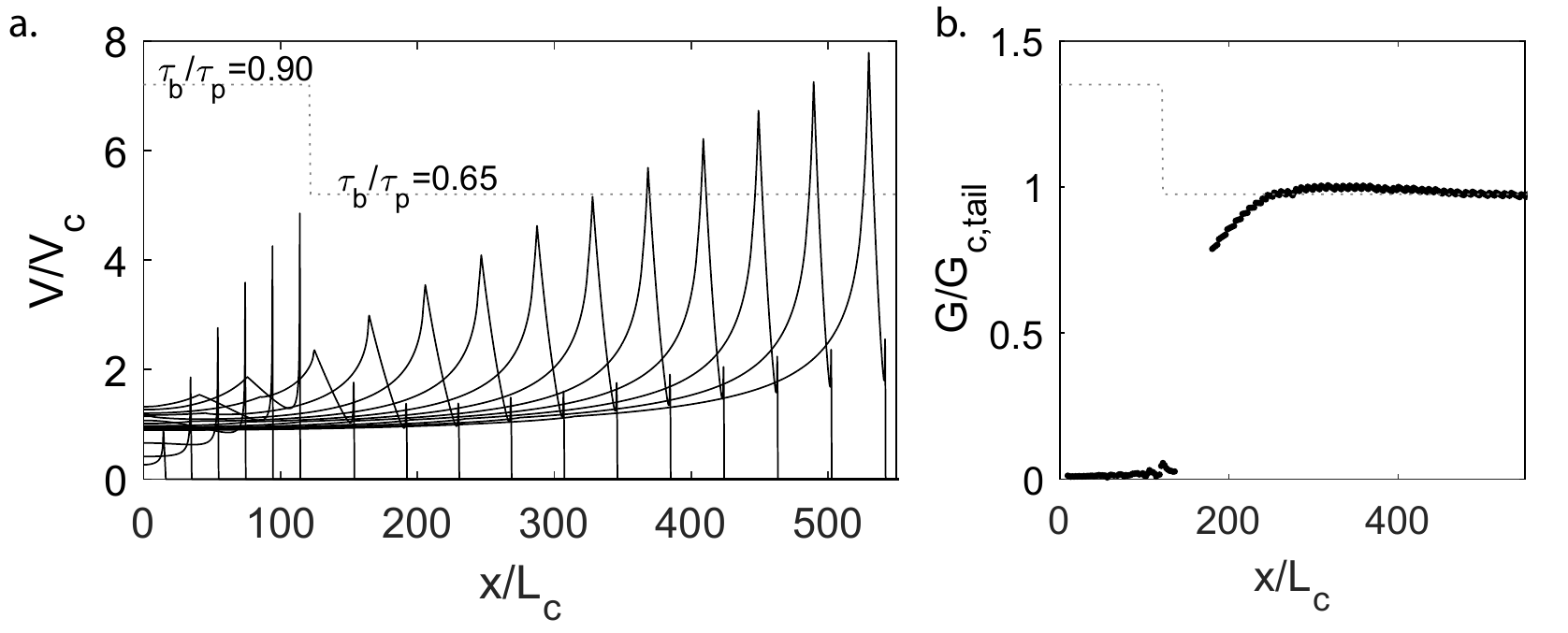}}
  \caption{a. Slip rate evolution with rupture length in presence of a stress barrier with rupture propagation controlled by the dual-scale weakening law. The initial background stress distribution is presented by the grey dashed line. b.  Energy release rate evolution with rupture length. Once overpassed the stress barrier, the energy release rate jumps to the value of fracture energy describing the long-tailed weakening $ G=G_\mathrm{c,tail} $ (i.e., rupture dynamics controlled by the long-tailed weakening).}
  \label{fig:5}
\end{figure}

We showed how the long-tailed weakening induces larger slip and higher slip velocities away from the crack tip. One consequence of this additional weakening is that it could help to overcome stress heterogeneities distributed along faults. 
To study this specific case, we impeded rupture acceleration by introducing a low stressed area at a distance $x/L_\mathrm{c}=120$ from the center of the fault, with $L_\mathrm{c}=\mu D\mathrm{c}/\tau_\mathrm{p}$.
The background stress, set initially at $\tau_\mathrm{b}/\tau_\mathrm{p}=0.90$ was decreased to $\tau_\mathrm{b}/\tau_\mathrm{p}=0.65$ in the outer region of the space domain. Under these conditions, once the rupture nucleates, it propagates generating two slip velocity peaks (Figure \ref{fig:5}a), in a similar way to the case without a stress barrier. However, due to the decrease of background stress, which is now much smaller than the residual stress associated to the first weakening $\tau_\mathrm{r} = 0.8\tau_\mathrm{p}$, the crack tip is momentarily stopped (since $ G < G_\mathrm{c,tip}$ ) at the location of the barrier. As time grows, the enhanced stress drop due to the prolonged weakening near the fault center is perceived by the crack tip, and promotes the propagation of the rupture across the barrier, which is observed as a second (large) peak slip rate taking over the rupture. The second weakening subsequently controls the complete rupture dynamics, following $G=G_\mathrm{c,tail}$ (Figure \ref{fig:5}b). These observations suggest that the large amount of slip induced by the long-tailed weakening allows the rupture to overcome zones of lower background stress that would normally stop the rupture controlled by the near-tip weakening only. \\

While a small value of energy is sufficient to nucleate and propagate a frictional rupture along fault interfaces, the presence of stress heterogeneities along a fault are expected to obstruct the propagation of ruptures induced by a rapid but limited frictional weakening. However, substantial weakening mechanisms activated at larger slip distances achieved in the central part of the crack can enhance the propagation of seismic rupture through regions of lower background stress, and control afterwards the dynamics of the crack. It emerges a possible scale dependence in the dynamics of rupture controlled by multiple weakening stages, meaning that critical cracks presenting large values of fracture energy can propagate due to the activation of slip on smaller cracks, which present lower values of fracture energy (i.e. enhancing propagation). This seems in agreement with recent experimental results highlighting that frictional instabilities are initiated by small events growing and cascading up into a much larger rupture \citep{mclaskey2014preslip}. This cascade of weakening mechanisms is also consistent with the sequence of deformation processes reported in fields observation of exhumed fault zones \citep{Incel2019}. Following our interpretation, the origin of breakdown work inverted from seismological observations could be related to energy dissipated through frictional weakening mechanisms rather than to the one dissipated near-edge (i.e. fracture energy of the interface). In fact, while the onset of friction is described by standard fracture processes, as stated in previous studies \citep{Svetlizky2014}, earthquake motions could be related to frictional weakening processes at the scale of crustal faults, which are expected to promote large values of breakdown work due to the activation of thermal processes during seismic slip \citep{DiToro2011}, and to present a clear dependence with slip, as observed for natural earthquakes \citep{Abercrombie2005, nielsen2016g}.

\section{Conclusions and relevance for natural observations}
Our results presented above highlight that:

i) A two-stage fault weakening is observed experimentally during frictional rupture propagation. A first rapid decay occurs within few microns of slip (ascribed to the critical slip distance $ D_\mathrm{c} $), followed by a long-tailed weakening, for which a steady state residual strength is not achieved at the scale of our experiments.

ii) The energy dissipated at the rupture tip is associated with the first weakening stage, defined here as the fracture energy of the interface $ G_\mathrm{c} $. This energy is the one controlling the onset of frictional rupture as already shown \citep{Svetlizky2014}. The energy dissipated during the long-tailed weakening corresponds to the breakdown work, which describes frictional weakening processes occurring at the interface during seismic slip.

iii) The derivation of the energy balance through the analysis of the stress intensity factors shows that further weakening, occurring once fracture energy is dissipated, will produce an additional energy release. This is expected to grow with time as more and more slip is achieved, enhancing the energy release rate at the crack tip and facilitating rupture propagation.

iv) Numerical simulations reveal the interplay between two successive weakening mechanisms represented by a dual-scale slip weakening law. The rapid near-tip weakening mechanism controls the propagation dynamics in regions of high background stress ($\tau_\mathrm{b}>\tau^*_\mathrm{b}$) where rupture is expected to nucleate. Once the nucleated rupture has generated sufficient slip to activate the second weakening mechanism, the resulting long-tail dissipation is able to drive the rupture further into portions of the fault with lower background stress ($\tau_\mathrm{b}<\tau^*_\mathrm{b}$) and across stress barriers.\\

\begin{figure}
\centerline{\includegraphics[width=1.2\linewidth]{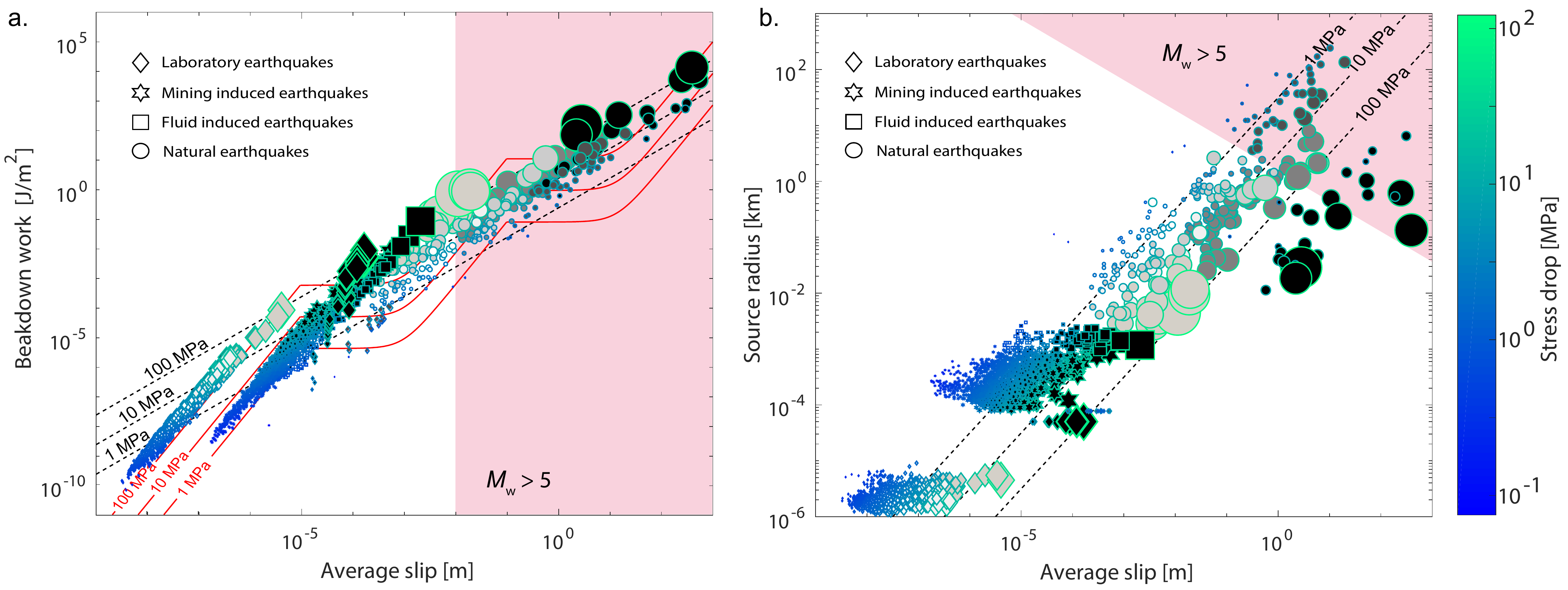}}
  \caption{a. Scale dependence of the effective fracture energy with the size of the seismic rupture, here represented by the size of the seismic slip. Diamonds, hexagrams, squares, circles correspond to breakdown work inverted in laboratory earthquakes, mining earthquakes, fluid-induced earthquakes, and natural earthquakes, respectively. Colors differentiate the population of events occuring along a same experimental setup, same mines or injection sites, and a same fault zone area.  The black dashed lines correspond to the evolution of the breakdown work as a function of the average slip assuming source model in infinite medium \citep{madariaga1976dynamics}, for three different stress drops. The red lines correspond to the prediction using  the triple slip weakening constitutive law described in the manuscript for three different values of initial shear stress. b. Scaling relation between the average slip and the source radius for the same populations of events. Dashed black lines  corresponds to the linear evolution of slip with rupture length assuming different stress drops. The color bar and the size of the symbols in (a.) and (b.) correspond to the stress drop estimated for each event. References to the data plot in this figure can be found in Supplementary Material.}
  \label{scaling}
\end{figure}

Remarkably, the scaling relationships between seismic slip and breakdown work inverted for mining, induced seismicity, laboratory earthquakes and natural earthquakes also shed light on possible scale-dependent breakdown work. 
At first sight, the breakdown work of natural earthquakes appears to increase linearly with seismic slip  \citep{Abercrombie2005,Tinti2005,nielsen2016g,selvadurai2019laboratory}. Such behavior is expected from earthquakes scaling laws which imply a stress drop independent of the earthquake size, and only a function of the ratio between the slip and the rupture length. For instance, assuming a circular crack model \citep{Brune1970,madariaga1976dynamics}, theoretical predictions of the breakdown work as a function of slip can be made for a given stress drop ($\Delta\sigma$) (Figure \ref{scaling}a), following $G_\mathrm{c}\approx\frac{1}{4}\Delta\sigma \bar{D}$ \citep{madariaga2009}, with $ \bar{D} $ the average slip. This model predicts a linear dependence between the breakdown work and the average slip, and agrees with breakdown work values estimated for the large earthquakes (with moment magnitude $M_w>5$) observed in nature (Figure \ref{scaling}a). However (taking the published breakdown work estimates at face value) conversely to theoretical predictions, each subset of smaller earthquakes ($M_w<5$ in either experimental faults, mines, injections sites or groups within the same fault zone) seems to follow independent power law (with exponent 2) relation (Figure \ref{scaling}a). This observation is compatible with a linear slip weakening behavior and is explained by the fact that for these events the average slip is not only a function of the rupture length, but it also increases with the stress drop for similar rupture lengths (Figure \ref{scaling}b). This suggests that, conversely to a circular crack model that considers ruptures propagating in an infinite medium (inducing a linear increase of slip with rupture length (Figure \ref{scaling}b)), the seismic rupture propagating during these earthquakes might be finite or geometrically constrained at boundaries, inducing a larger release of stress through an increase of slip. 

Could then earthquakes obey to a single slip dependent constitutive law through different length scales? This was proposed by \citet{Viesca2015}, who showed that a transition from flash heating to thermal pressurization could explain a wide range of observations. Such behavior would imply a continuous increase of the stress drop with slip, which would not be compatible with seismological observations highlighting that earthquakes of much different magnitudes present similar values of stress drops (Figures \ref{scaling}). However, the stress drop is a function of the constitutive law (and for a same constitutive law, of the final slip), but also a function of the initial shear stress acting on the fault, which might differ spatially along the fault itself and change with focal depth. Here, we explore how a multiple-scale slip weakening law is compatible with seismological and experimental estimates of breakdown work. We computed the evolution of the breakdown work as a function of slip for different values of initial shear stress (1, 10, 100 MPa), using a triple slip weakening constitutive law. This simplified frictional constitutive law was established assuming three distinct weakening mechanisms that are known to operate at different length scales and to present different critical slip distances: (i) flash heating phenomena (heating of fault asperities due to shear) which induces large strength release within the first microns of slip ($\Delta f =0.4, D_c= 100 \mu$m) \citep{rice2006heating,goldsby2011flash}, (ii) melt lubrication (melting of the bulk due to shear heating) or thermal pressurization (heating and expansion of fluids which weaken the shear zone) which are activated at intermediate slip distances ($\Delta f = 0.4$ over a slip distance of $ D_c= 0.1$ m) \citep{hirose2005growth,rice2006heating,DiToro2011}, (iii) thermal decomposition of the fault's minerals which induces a slight decrease in fault strength at large slip ($\Delta f =0.2$ for slip ranging from  0.1 to 100 meters) \citep{han2007ultralow,sulem2009thermal,Brantut2010} (Figure \ref{scaling}a). Computing the breakdown work as a function of slip for this constitutive law highlights that each different stage of weakening presents a distinct scaling relation (i.e. power-law with an exponent 2) at the different length scales, allowing to jump from one population of events to the other (Figure \ref{scaling}a). The scaling observed for natural earthquakes is well described by this triple slip weakening constitutive law, suggesting that earthquakes spanning all possible range of magnitudes could obey to a similar set of constitutive laws across the different length scales, and that the amount of breakdown work resulting from rupture propagation is a result of the final slip and of the initial shear stress acting along the fault.

Our approach is naturally very simplified and has some limitations. In nature, weakening mechanisms are not expected to follow a linear slip weakening behavior \citep[e.g.][]{Viesca2015}, which could modify the slip dependence of the breakdown work, as observed in recent studies \citep{Lambert2020}. However, the activation of different weakening mechanisms with increasing slip suggests that while the early stage of instabilities could be controlled mostly by fracture energy (i.e. the first weakening stage observed in our experiments), the complete breakdown work and energy release rate at the rupture tip is expected to increase with slip. In other words, while natural earthquakes might be expected to initiate like classical shear cracks, continued earthquake slip should be related to friction at the scale of the entire fault.

\section{Acknowledgments}
F.P., F.X.P. and M.V. would like to thank Jay Fineberg and Jean-François Molinari for helpful discussions. 
F.P., M.L. and M.V. acknowledge support by the European Research Council Starting Grant project 757290-BEFINE. F.B. acknowledges support of the Swiss National Science Foundation through the fellowship No. P2ELP2/188034. 

\bibliography{library2}

\end{document}